\documentclass[aps,prl,twocolumn,superscriptaddress,nofootinbib,longbibliography,floatfix]{revtex4-2}

\usepackage{amsmath,amssymb,bm}
\usepackage{graphicx}
\usepackage{microtype}
\usepackage{xcolor}
\usepackage{hyperref}
\hypersetup{hidelinks}

\newcommand{\Tr}{\operatorname{Tr}}
\newcommand{\cM}{\tilde{\mathcal{M}}^{(2)}}
\newcommand{\Mtwo}{\mathcal{M}^{(2)}}

\begin{document}

\title{Stabilizer-R\'enyi Microscopy of Critical Correlations in Interacting Fermions}

\author{Jun Qi Fang}
\affiliation{Key Laboratory of Artificial Structures and Quantum Control (Ministry of Education), School of Physics and Astronomy, Shanghai Jiao Tong University, Shanghai 200240, China}
\author{Fo-Hong Wang}
\affiliation{Key Laboratory of Artificial Structures and Quantum Control (Ministry of Education), School of Physics and Astronomy, Shanghai Jiao Tong University, Shanghai 200240, China}
\affiliation{Tsung-Dao Lee Institute, Shanghai Jiao Tong University, Shanghai 200240, China}
\author{Xiao Yan Xu}
\email{xiaoyanxu@sjtu.edu.cn}
\affiliation{Key Laboratory of Artificial Structures and Quantum Control (Ministry of Education), School of Physics and Astronomy, Shanghai Jiao Tong University, Shanghai 200240, China}
\affiliation{Hefei National Laboratory, Hefei 230088, China}

\begin{abstract}
Quantum magic---the resource that separates universal quantum computation from efficiently simulable Clifford circuits---has emerged as a diagnostic of many-body quantum states, yet the stabilizer R\'enyi entropy (SRE) that quantifies it remains largely inaccessible in interacting fermion systems.
Estimating the global SRE requires a sum over exponentially many Majorana strings, which in determinant quantum Monte Carlo can be sampled without a sign problem only for the interacting density matrix obtained after averaging the auxiliary fields: sampling strings and fields simultaneously incurs a sign problem even for sign-problem-free models, so that the known sign-free alternative is a nested Monte Carlo that does not scale.
We propose instead the two-point SRE---a practical stabilizer-R\'enyi correlator built from rank-2 SREs of one- and two-site reduced density matrices, which are fixed exactly by single-particle Green functions and density correlations already measured in standard simulations---and show that it obeys universal finite-size scaling at fermionic quantum critical points.
In the one-dimensional half-filled spinless $t$-$V$ chain, the correlator distinguishes algebraic and exponential decay regimes and tracks the inverse-logarithmic finite-size drift characteristic of the Berezinskii--Kosterlitz--Thouless transition.
On the honeycomb lattice, sign-problem-free quantum Monte Carlo yields Gross--Neveu--Ising scaling with finite anomalous dimension at zero temperature and two-dimensional Ising collapse at the thermal transition.
Our results suggest stabilizer-R\'enyi microscopy as a spatially resolved, quantitatively universal probe of critical correlations in interacting fermionic matter, on par with conventional order-parameter correlators and accessible to quantum-simulator measurements.
\end{abstract}

\maketitle

Entanglement has transformed how quantum matter is characterized and simulated
\cite{vidalEntanglementQuantumCritical2003,Amico2008rmp,calabreseEntanglementEntropyConformal2009,Laflorencie2016pr,eisertColloquiumAreaLaws2010,vidalEfficientClassicalSimulation2003}.
At one-dimensional quantum critical points it exhibits universal logarithmic scaling; in gapped phases it distinguishes short- and long-range entangled topological order; and its spatial structure explains why density matrix renormalization group (DMRG) succeeds when entanglement remains sufficiently constrained.
Entanglement is not, however, the sole resource controlling quantum computational complexity.
Clifford circuits generate highly entangled stabilizer states while remaining classically simulable via the Gottesman--Knill theorem
\cite{gottesmanHeisenbergRepresentationQuantum1998,aaronsonImprovedSimulationStabilizer2004}.
Universal quantum computation additionally requires non-Clifford gates that supply quantum magic, or nonstabilizerness
\cite{gottesmanDemonstratingViabilityUniversal1999,kitaevQuantumComputationsAlgorithms1997,bravyiUniversalQuantumComputation2005,reichardtQuantumUniversalityState2009,wangnpj2024}.
If entanglement distinguishes product states from correlated quantum matter, magic distinguishes the stabilizer manifold---which remains efficiently simulable despite arbitrary entanglement---from generic quantum states outside it.

This resource-theoretic distinction motivates magic as a complementary diagnostic of many-body physics
\cite{Veitch2012njp,Veitch2014njp,Howard2017prl,Pashayan2015prl,StabilizerRenyiEntropy,spriggsQuantumResourcesQuantum2025}.
Stabilizer R\'enyi entropies (SREs) recast magic in terms of moments of Pauli-string expectation values, making it computable for many-body states
\cite{StabilizerRenyiEntropy,leoneStabilizerEntropiesAre2024,lamiQuantumMagicPerfect2023,haugQuantifyingNonstabilizernessMatrix2023}.
Numerical studies of quantum spin models have since mapped the critical behavior of magic and the magic stored in correlations between subsystems
\cite{lamiQuantumMagicPerfect2023,tarabungaManyBodyMagicPauliMarkov2023,frauNonstabilizernessEntanglementMatrix2024,tarabungaCriticalBehaviorsNonstabilizerness2024,dingEvaluatingManybodyStabilizer2025,timsinaRobustnessNonstabilizernessQuantum2025}.
For fermions, however, progress has centered on Gaussian states and exactly solvable Sachdev--Ye--Kitaev models
\cite{colluraQuantumMagicFermionic2025,beraNonStabilizernessSachdevYeKitaevModel2025,zhangStabilizerRenyiEntropy2025,iannottiNonLocalMagicResources2026}.
The obstruction in a generic interacting system is elementary: after Jordan--Wigner transformation, a global fermionic SRE contains exponentially many Majorana-string correlators, and determinant quantum Monte Carlo (DQMC) represents an interacting density matrix as an ensemble of Gaussian operators.
The global SRE is exponentially small in the system size, so the string sum must be importance sampled; configuration-resolved amplitudes fluctuate in sign, however, so a non-negative sampling weight exists only for even powers of the field-averaged amplitude, and in this representation the sign-free option is a nested string--field sampling whose signal-to-noise ratio degrades rapidly with system size [Supplemental Material (SM)].

We replace the global question with a local one: how does the SRE functional respond to correlations between two spatial points?
The resulting stabilizer-R\'enyi correlator uses only the two-site reduced density matrix, which for number-conserving spinless fermions is reconstructed exactly from one-body Green functions, local densities, and a density correlator---standard outputs of DMRG and DQMC, and measurable in ultracold-atom quantum simulators
\cite{Bloch2012quantum,parezEntanglementNegativitySeparated2024}.
No high-order string sampling is required.
We validate the construction at the exactly known one-dimensional Berezinskii--Kosterlitz--Thouless (BKT) transition and then deploy sign-problem-free DQMC to resolve quantitative universal scaling at a two-dimensional fermionic quantum critical point and its finite-temperature counterpart.
The central finding is that the stabilizer R\'enyi functional, evaluated on two-site marginals, is a spatially resolved observable obeying the same finite-size-scaling and universality-class identification tests as conventional many-body correlators.

For a density matrix $\rho_A$ on $N_A$ fermionic modes, we use the raw mixed-state extension of the rank-2 SRE,
\begin{equation}
 \Mtwo(\rho_A)
 =-\ln\!\left[\frac{1}{2^{N_A}}\sum_{P_A}
       \left|\Tr(P_A\rho_A)\right|^4\right]
   +\ln\Tr\rho_A^2 ,
\label{eq:M2}
\end{equation}
where $P_A$ denotes the Pauli strings associated with a fixed ordering of the modes.
We isolate the correlation-dependent part by defining the two-point SRE, or stabilizer-R\'enyi correlator,
\begin{equation}
 \cM_{i,j}=\Mtwo(\rho_{ij})-\Mtwo(\rho_i)-\Mtwo(\rho_j),
\label{eq:CM}
\end{equation}
where the tilde marks the connected two-point quantity.
Subtracting the two single-site terms removes the local background and makes $\cM$ vanish when $\rho_{ij}=\rho_i\otimes\rho_j$.
Like the SRE itself, it depends on the chosen local mode basis; throughout this work the basis is the physical lattice-site basis.
The qualifier ``correlator'' is deliberate: unlike the convex-roof extension that defines a mixed-state monotone (SM), the raw functional in Eq.~(\ref{eq:M2}) lacks a partial-trace monotonicity guarantee, and $\cM$ need not be positive
\cite{leoneStabilizerEntropiesAre2024,haugEfficientWitnessingTesting2026}.
We use it as a connected SRE diagnostic of spatial correlation structure rather than as a magic measure.

Fermion parity and particle-number conservation restrict the two-site reduced density matrix to five independent parameters.
In the ordered basis
$\{|0\rangle,c_j^\dagger|0\rangle,c_i^\dagger|0\rangle,
c_i^\dagger c_j^\dagger|0\rangle\}$,
\begin{equation}
\rho_{ij}=
\begin{pmatrix}
1-n_i-n_j+n_{ij}&0&0&0\\
0&n_j-n_{ij}&g_{ij}&0\\
0&\overline{g}_{ij}&n_i-n_{ij}&0\\
0&0&0&n_{ij}
\end{pmatrix},
\label{eq:rhoij}
\end{equation}
where $n_i=\langle n_i\rangle$, $n_{ij}=\langle n_i n_j\rangle$, and
$g_{ij}=\langle c_i^\dagger c_j\rangle$.
Two local densities, one complex Green function, and one density correlator form a complete tomography set
\cite{parezEntanglementNegativitySeparated2024}.
Equation~(\ref{eq:rhoij}) makes $\cM$ an exact post-processing observable: the correlators are read directly from DMRG, or averaged over the auxiliary-field ensemble in DQMC before the nonlinear SRE is evaluated.

At half filling, unbroken particle--hole symmetry sets $n_i=n_j=1/2$.
Writing $\delta_{ij}=\langle n_i n_j\rangle-\langle n_i\rangle\langle n_j\rangle$ and $g_{ij}=g_R+\text{i} g_I$, with $g_R$ and $g_I$ the real and imaginary parts, the result is
\begin{equation}
 \cM_{i,j}=
 \ln\frac{1+8|g_{ij}|^2+16\delta_{ij}^2}
 {1+32(g_R^4+g_I^4)+256\delta_{ij}^4}.
\label{eq:exactCM}
\end{equation}
Its long-distance expansion is
\begin{equation}
 \cM_{i,j}=8|g_{ij}|^2+16\delta_{ij}^2+
 O(g_{ij}^4,g_{ij}^2\delta_{ij}^2,\delta_{ij}^4).
\label{eq:expansion}
\end{equation}
Equations~(\ref{eq:exactCM}) and (\ref{eq:expansion}) reduce the stabilizer-R\'enyi correlator to two inputs that enter quadratically at leading order: the Green function $g_{ij}$ and the connected density correlator $\delta_{ij}$.
An algebraically decaying input therefore yields an algebraic $\cM$ with doubled scaling dimension, and an exponentially decaying input yields exponential decay; at long distance, $\cM$ follows whichever of the two inputs decays more slowly.

\begin{figure*}[t]
\centering
\includegraphics[width=\textwidth]{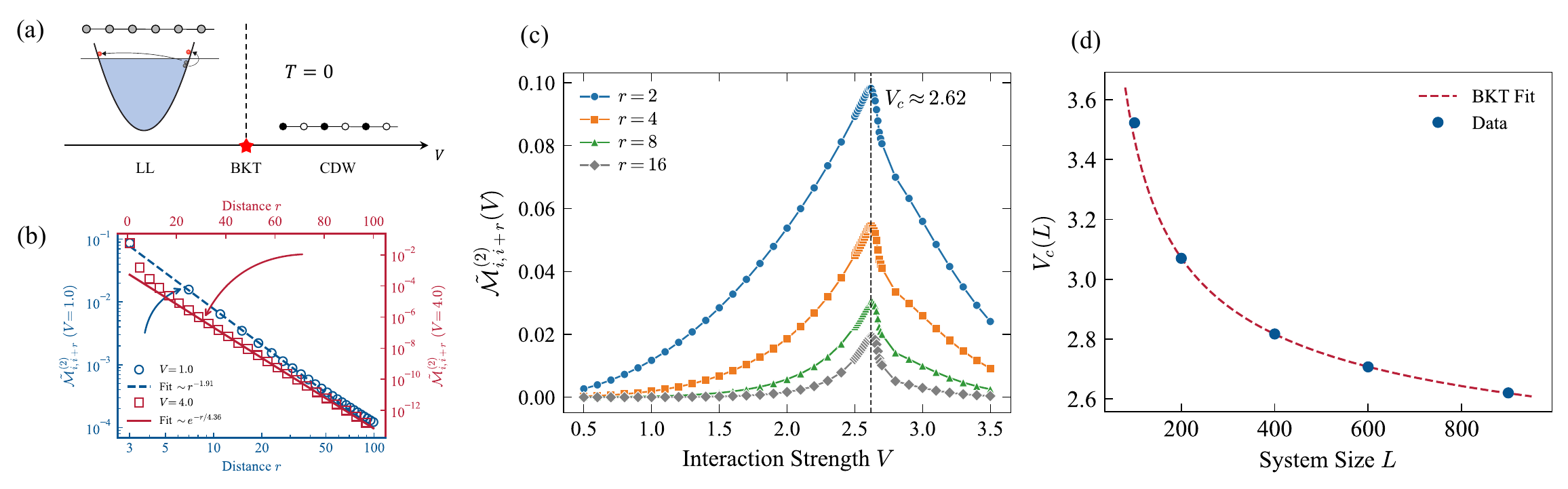}
\caption{\textbf{Stabilizer-R\'enyi correlations resolve the one-dimensional BKT transition.}
(a) Schematic phase diagram of the half-filled spinless $t$-$V$ chain across the BKT transition. (b) DMRG results for $\cM_{i,i+r}$ on an open $L=400$ chain.
The Luttinger-liquid point $V=1$ shows algebraic decay (fit $r^{-1.91}$), whereas the CDW point $V=4$ shows exponential decay (fit $\exp[-r/4.36]$).
(c) Interaction dependence for $L=900$ at four separations.
The maximum defines the size-dependent pseudocritical coupling $V_c(L)$.
(d) Peak positions approach the exact $V_c=2$ according to
$V_c(L)=2+\pi^4/[4(\ln L-\ln L_0)^2]$, with $L_0\simeq1.7$ for $L>100$.}
\label{fig:1D}
\end{figure*}

We study the half-filled nearest-neighbor model
\begin{equation}
H=-t\sum_{\langle i,j\rangle}
  (c_i^\dagger c_j+c_j^\dagger c_i)
 +V\sum_{\langle i,j\rangle}
  \left(n_i-\frac12\right)\left(n_j-\frac12\right),
\label{eq:tV}
\end{equation}
and set $t=1$.
On a chain, Eq.~(\ref{eq:tV}) has a Luttinger-liquid phase for $V<2$ and a charge-density-wave (CDW) phase for $V>2$, separated by a BKT transition [Fig.~\ref{fig:1D}(a)]
\cite{affleckCriticalBehaviourSpins1989,cazalillaOneDimensionalBosons2011,mishraPhaseDiagramHalffilled2011}.
We use matrix-product-state DMRG with open boundaries and choose the reference site near the center to suppress boundary effects
\cite{whiteDensityMatrixFormulation1992,itensor,itensor-r0.3}.
The thermodynamic transition is exactly fixed at $V_c=2$, but the exponentially divergent BKT correlation length produces logarithmically slow finite-size drift---a stringent test that requires the diagnostic to reproduce more than the distinction between phases.

Figure~\ref{fig:1D}(b) shows the distance dependence on an $L=400$ chain.
At $V=1$ the data follow the algebraic fit $\cM_{i,i+r}\propto r^{-1.91}$, while at $V=4$ they follow $\cM_{i,i+r}\propto\exp(-r/4.36)$.
These are empirical fits to the stabilizer-R\'enyi correlator; their significance is the qualitative change of spatial structure.
In the Luttinger liquid, both inputs to Eq.~(\ref{eq:exactCM}) decay algebraically and the leading quadratic terms retain a long-ranged tail.
In the gapped CDW regime the DMRG ground state breaks particle--hole symmetry spontaneously, so the densities are staggered and we evaluate the general expressions of the SM.
Their leading term is linear in $\delta_{ij}$, and the stabilizer-R\'enyi correlator therefore inherits the density correlation length itself.

At fixed separation, $\cM$ develops a pronounced maximum as $V$ is varied [Fig.~\ref{fig:1D}(c)].
The maximum reflects the competition between correlations enhanced on approaching the transition and their exponential suppression after entering the gapped regime.
For a finite chain the peak occurs at a pseudocritical coupling $V_c(L)>2$, where the BKT correlation length becomes comparable to $L$; its conspicuous displacement even on hundreds of sites is the characteristic BKT finite-size effect.

To test this quantitatively, we parameterize the ordered side by $V/2=\cosh\lambda$.
Close to the transition, $\lambda\simeq\sqrt{V-2}$ and the leading Bethe-ansatz scale is
$\xi=\xi_0\exp[\pi^2/(2\lambda)]$.
Equating $\xi$ and $L$ gives
\begin{equation}
 V_c(L)=2+\frac{\pi^4}{4[\ln L-\ln L_0]^2},
\label{eq:BKTdrift}
\end{equation}
where $L_0$ absorbs the microscopic prefactor and subleading corrections.
The form is the slow inverse-logarithmic drift familiar from BKT transitions
\cite{okamotoFluiddimerCriticalPoint1992,eggertNumericalEvidenceMultiplicative1996,laflorencieFiniteSizeScaling2001,sunFidelityBerezinskiiKosterlitzThoulessQuantum2015,liEntanglementEntropyBerezinskii2016,spaldingCorrectionsConformalCharge2024}.
It describes the peak positions through $L=900$; fitting $L>100$ gives $L_0\simeq1.7$ [Fig.~\ref{fig:1D}(d)].
The 1D calculation benchmarks three levels: algebraic versus exponential spatial decay, a finite-size peak near the phase boundary, and the unusually slow universal flow of that peak toward the exact transition.

\begin{figure*}[t]
\centering
\includegraphics[width=\textwidth]{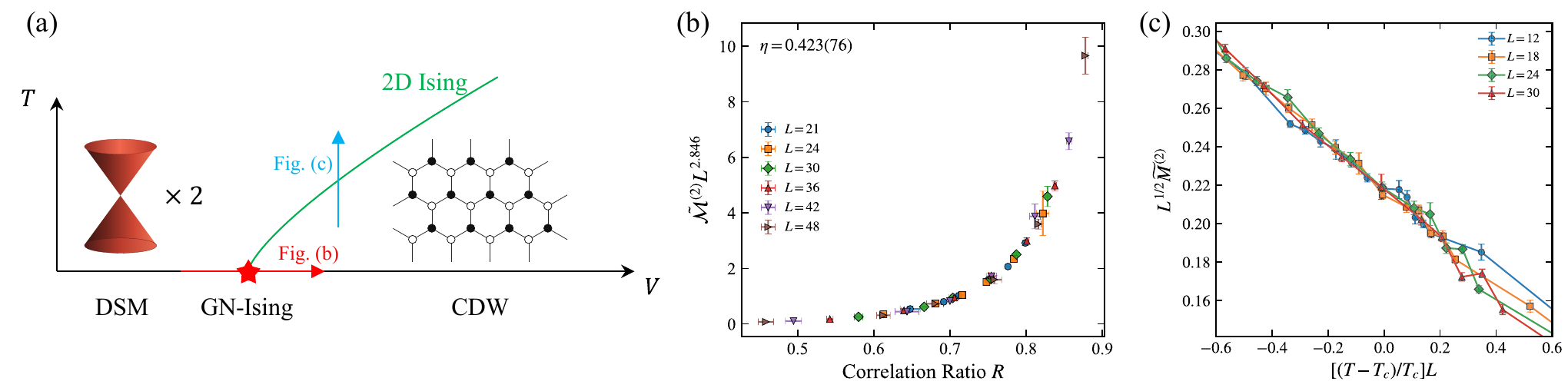}
\caption{\textbf{QMC stabilizer-R\'enyi microscopy at fermionic critical points.}
(a) Schematic phase diagram of the half-filled spinless $t$-$V$ model on the honeycomb lattice in the interaction--temperature plane.
(b) Scaling collapse against the dimensionless CDW correlation ratio
$R=1-C(\bm q_{\min})/C(\bm 0)$.
The collapse of $L^{2+2\eta}\cM$ gives $\eta=0.423(76)$; couplings $V>1.36$ are excluded.
(c) Thermal collapse of $L^{1/2}\cM$ versus $[(T-T_c)/T_c]L$ at $V=2.0$ for $L=12,18,24,30$, with powers fixed to two-dimensional Ising values.
Error bars show DQMC statistical uncertainties.}
\label{fig:honeycomb}
\end{figure*}

The decisive test is an interacting fermionic critical point in two dimensions.
On the honeycomb lattice, Eq.~(\ref{eq:tV}) undergoes a transition from a Dirac semimetal to a CDW insulator near $V_c\simeq1.34$, in the Gross--Neveu--Ising universality class [Fig.~\ref{fig:honeycomb}(a)]
\cite{wangFermionicQuantumCritical2014,liFermionsignfreeMajaranaquantumMonteCarloStudies2015,Hesselmann2016,Wang2016,wangfohong2026}.
This transition combines gapless Dirac quasiparticles with a fluctuating Ising order parameter.
At half filling the nearest-neighbor model admits a sign-problem-free formulation
\cite{Huffman2014,Li2015solving,Wang2015split}.
We use DQMC \cite{blankenbeclerMonteCarloCalculations1981,scalapinoMonteCarloCalculations1981,assaadWorldlineDeterminantalQuantum2008} to measure the equal-time Green function and density correlations, reconstruct Eq.~(\ref{eq:rhoij}), and evaluate $\cM$ between a maximally separated pair, whose displacement is $\bm r_{\max}$.

At a Lorentz-invariant critical point ($d=2$, dynamical exponent $z=1$), the maximally separated CDW correlator obeys
\begin{equation}
C_{\max}(L)=
\left\langle(n_i-\tfrac12)(n_{i+\bm r_{\max}}-\tfrac12)\right\rangle
\sim L^{-(1+\eta)} .
\end{equation}
The fermion Green function decays as
$g_{i,i+\bm r_{\max}}\sim L^{-(2+\eta_\psi)}$, with $\eta_\psi\geq0$ by unitarity, so the density term in Eq.~(\ref{eq:expansion}) decays more slowly whenever $\eta<1$.
This holds throughout the Gross--Neveu--Yukawa class, where $\eta=1$ is the $N\to\infty$ value fixed by Dirac Landau damping and $1/N$ corrections lower it, leaving $\eta\simeq0.5$ and $\eta_\psi\simeq0.1$ at the present fixed point (SM)
\cite{Phys.Rev.B2018Ihrig,graceyAnomalousDimensionsCritical2025,Phys.Lett.B1992Gracey,J.HighEnerg.Phys.2023Erramilli}.
The density channel is therefore asymptotically dominant, yielding
\begin{equation}
\cM_{i,i+\bm r_{\max}}\sim L^{-2(1+\eta)} .
\label{eq:CM_scaling}
\end{equation}
The doubled exponent is a parameter-free consequence of the exact tomography formula.

To reduce sensitivity to the critical coupling, we use the dimensionless correlation ratio
$R=1-C(\bm q_{\min})/C(\bm0)$, with
$\bm q_{\min}=(2\pi/L,0)$, as the horizontal coordinate
\cite{haradaKernelMethodCorrections2015,ChiralHeisenbergGrossNeveuYukawa}.
Unlike a direct plot against $(V-V_c)L^{1/\nu}$, this does not require simultaneous determination of $V_c$ and $\nu$ from the correlator data.
Near criticality, $R$ depends on $V$ and $L$ only through $(V-V_c)L^{1/\nu}$ up to corrections to scaling, so it serves as a dimensionless scaling coordinate.
The raw longest-distance data, shown in the SM, distinguish the semimetallic and ordered trends: around $V=1.34$ the signal follows a power law over the available sizes, while departures become visible on either side.

For $L=21,24,30,36,42,48$ and $V\leq1.36$, the data collapse as
$L^{2+2\eta}\cM=\mathcal F(R)$ with $\eta=0.423(76)$
[Fig.~\ref{fig:honeycomb}(b)].
We exclude $V>1.36$, where ordered-side corrections are visibly larger.
The extracted $\eta$ agrees with independent estimates for the Gross--Neveu--Ising fixed point
\cite{Phys.Rev.B2018Ihrig,graceyAnomalousDimensionsCritical2025,Phys.Lett.B1992Gracey,Int.J.Mod.Phys.A1994Graceya,Int.J.Mod.Phys.A1994Gracey,Phys.Rev.B2016Knorr,J.HighEnerg.Phys.2023Erramilli,Sci.Adv.2026YK.Yu,liFermionsignfreeMajaranaquantumMonteCarloStudies2015,Phys.Rev.D2020Huffman,wangfohong2026}
and, importantly, is obtained from an SRE-based observable with DQMC statistical uncertainties.
The result is not a numerical coincidence: the factor of two in the scaling dimension follows from Eq.~(\ref{eq:exactCM}) and is borne out over the critical window.
It demonstrates that the local construction preserves quantitative universality in a genuinely interacting two-dimensional fermion calculation.

The same estimator remains well defined for a thermal density matrix.
The finite-temperature CDW transition [Fig.~\ref{fig:honeycomb}(a)] breaks discrete sublattice symmetry and has two-dimensional Ising exponents $\eta_{\rm I}=1/4$ and $\nu_{\rm I}=1$.
Equation~(\ref{eq:expansion}) predicts
\begin{equation}
\cM(T,L)=L^{-2\eta_{\rm I}}\,
\mathcal G\!\left(\frac{T-T_c}{T_c}L^{1/\nu_{\rm I}}\right).
\label{eq:thermal}
\end{equation}
The prefactor $L^{-2\eta_{\rm I}}=L^{-1/2}$ again follows because $\cM$ is led by the square of the order-parameter correlator.
At $V=2.0$ in the CDW regime, DQMC data for $L=12,18,24,30$ collapse with the fixed Ising powers without changing the estimator [Fig.~\ref{fig:honeycomb}(c)].
This fixed-exponent collapse contrasts with the global rank-2 SRE, which was reported to lose sensitivity at finite-temperature transitions \cite{dingEvaluatingManybodyStabilizer2025}, and shows that this limitation is not generic to SRE-derived thermal diagnostics.
For detecting thermal criticality, rather than quantifying total magic, the two-point SRE can therefore be more informative than its global counterpart.
Together, the quantum and thermal results show that the same tomography protocol follows two different universality classes without tailoring the observable to either transition.

\textit{Discussion.---}
Our results establish a scalable notion of stabilizer-R\'enyi microscopy built on the two-point SRE: an exact interface between SREs and the observables that interacting-fermion methods already calculate.
The same reduced-density-matrix estimator follows the inverse-logarithmic BKT drift in one dimension and Gross--Neveu--Ising and two-dimensional Ising criticality in two, across ground-state DMRG, zero-temperature projector DQMC, and finite-temperature DQMC.

Equation~(\ref{eq:exactCM}) shows that $\cM$ is a nonlinear combination of the Green function and the connected density correlator; it cannot contain more information than the two-site density matrix from which it is constructed.
The exponent $2(1+\eta)$ at the honeycomb transition is fixed by the CDW anomalous dimension, not an independent critical exponent.
What the SRE contributes is the combination itself: its form follows from Eq.~(\ref{eq:M2}), and the single-site subtraction makes $\cM$ vanish for uncorrelated sites, so a nonzero value measures how correlations shift the two-site Pauli-moment content away from the sum of its single-site values.

The two-site restriction that limits the observable also makes it tractable.
The global rank-2 SRE obeys a volume law \cite{dingEvaluatingManybodyStabilizer2025}, but estimating it requires nested string--field sampling---forced by the joint-sampling sign problem---whose signal degrades rapidly with size.
Restricting the subsystem before evaluating the SRE reduces the exponential string sum to the 16 Pauli operators of two modes.
The many-body cost is then only that of obtaining the local densities, $g_{ij}$, and $\langle n_i n_j\rangle$.
This tradeoff sacrifices knowledge of the total magic content, but it makes interacting systems, finite temperatures, and lattice sizes beyond global SRE sampling accessible with no new QMC estimator.

The observable is local, basis dependent, and constructed from a raw mixed-state SRE; it is not a convex-roof resource monotone and does not measure the total classical simulation cost.
The two-site marginals studied here are in fact stabilizer mixtures, hence free states carrying exactly zero magic (SM).
Since the global SRE nonetheless obeys a volume law, the magic of these fermionic states is intrinsically nonlocal: no two-site reduced density matrix can hold any of it, and $\cM$ must be read as a Pauli-moment correlation functional rather than a local magic density.
Nor can a two-site quantity diagnose intrinsically nonlocal order.
The SM treats a Laughlin-orbital calculation as exploratory: in the orbital basis on the sphere, $\cM$ resolves the short-range exclusion window and its filling-controlled plateau, but that plateau is neither basis-independent nor evidence by itself for topological order.
Establishing genuinely nonlocal fermionic magic will require larger reduced density matrices or optimized nonlocal measures
\cite{qianQuantumNonlocalNonstabilizerness2025,andreadakisExactLinkNonlocal2026,iannottiNonLocalMagicResources2026}.

The same limitations point to concrete extensions.
Three- and four-site tomography could locate the smallest cluster whose reduced density matrix carries certifiable magic, while clusters of increasing diameter could interpolate, at growing tomography cost, between the present microscope and a global SRE.
Spinful and multiorbital models require larger local density matrices but retain the basic strategy whenever symmetry reduces the tomography set.
A natural next target is the half-filled honeycomb Hubbard model, where sign-problem-free DQMC could test the same construction at Gross--Neveu--Heisenberg criticality, with spin correlations replacing the CDW channel
\cite{ChiralHeisenbergGrossNeveuYukawa}.
Because the required one- and two-body correlators are standard simulation outputs and accessible to local tomography protocols, the construction offers a route for comparing classical many-body calculations with quantum simulators
\cite{Bloch2012quantum,parezEntanglementNegativitySeparated2024}.

\begin{acknowledgments}
We thank Cheng-Hao He for fruitful discussions.
This work was supported by the National Natural Science Foundation of China (Grants No.~12447103, No.~12274289, and No.~125B2077), the National Key R\&D Program of China (Grants No.~2022YFA1402702 and No.~2021YFA1401400), the Quantum Science and Technology-National Science and Technology Major Project (Grant No.~2021ZD0301902), the Yangyang Development Fund, and the Shanghai Jiao Tong University 2030 Initiative.
F.-H.W. is supported by the T.~D. Lee Scholarship.
The computations were performed on the Siyuan-1 and $\pi$ 2.0 clusters at the Center for High Performance Computing, Shanghai Jiao Tong University.
\end{acknowledgments}

\bibliography{ref}

\end{document}


\title{Supplemental Material for ``Stabilizer-R\'enyi Microscopy of Critical Correlations in Interacting Fermions''}

\author{Jun Qi Fang}
\affiliation{Key Laboratory of Artificial Structures and Quantum Control (Ministry of Education), School of Physics and Astronomy, Shanghai Jiao Tong University, Shanghai 200240, China}
\author{Fo-Hong Wang}
\affiliation{Key Laboratory of Artificial Structures and Quantum Control (Ministry of Education), School of Physics and Astronomy, Shanghai Jiao Tong University, Shanghai 200240, China}
\affiliation{Tsung-Dao Lee Institute, Shanghai Jiao Tong University, Shanghai 200240, China}
\author{Xiao Yan Xu}
\email{xiaoyanxu@sjtu.edu.cn}
\affiliation{Key Laboratory of Artificial Structures and Quantum Control (Ministry of Education), School of Physics and Astronomy, Shanghai Jiao Tong University, Shanghai 200240, China}
\affiliation{Hefei National Laboratory, Hefei 230088, China}

\maketitle

This Supplemental Material contains: (I) the definition and scope of the mixed-state stabilizer R\'enyi functional used in the Letter; (II) the exact two-site tomography formula, its correlation expansion, and a proof that the two-site marginals studied here are stabilizer mixtures; (III) numerical and scaling details for the one-dimensional and honeycomb-lattice $t$-$V$ models; (IV) the global-SRE estimator, the sign-problem origin of its nested sampling, and the resulting numerical bottleneck; and (V) an exploratory Laughlin-orbital calculation.

\section{Rank-2 stabilizer R\'enyi functional and its scope}

For a density matrix $\rho_A$ acting on $N_A$ modes, we define
\begin{equation}
\mathcal M^{(2)}(\rho_A)
=-\ln\!\left[
\frac{1}{2^{N_A}}\sum_{P_A\in\mathcal P_{N_A}}
\left|\Tr(P_A\rho_A)\right|^4
\right]+\ln\Tr\rho_A^2 .
\label{eq:SRE2}
\end{equation}
Using Pauli-basis completeness,
$\Tr\rho_A^2=2^{-N_A}\sum_{P_A}|\Tr(P_A\rho_A)|^2$, Eq.~(\ref{eq:SRE2}) is equivalently
\begin{equation}
\mathcal M^{(2)}(\rho_A)
=\ln\frac{\sum_{P_A}|\Tr(P_A\rho_A)|^2}
{\sum_{P_A}|\Tr(P_A\rho_A)|^4}.
\label{eq:SRE2ratio}
\end{equation}
For sites $i$ and $j$, the connected quantity used throughout the Letter is
\begin{equation}
\cM_{i,j}=
\mathcal M^{(2)}(\rho_{ij})
-\mathcal M^{(2)}(\rho_i)
-\mathcal M^{(2)}(\rho_j).
\label{eq:connected}
\end{equation}

The original SRE is a pure-state magic measure
\cite{StabilizerRenyiEntropy}, and pure-state monotonicity for R\'enyi index $\alpha\geq2$ has been established
\cite{leoneStabilizerEntropiesAre2024}.
For mixed states, a genuine monotone is obtained by the convex-roof extension
\begin{equation}
\mathcal M^{(2)}_{\mathrm{cr}}(\rho)
=\min\sum_k p_k\,\mathcal M^{(2)}(|\psi_k\rangle),
\label{eq:convexroof}
\end{equation}
the minimum running over all decompositions
$\rho=\sum_k p_k|\psi_k\rangle\langle\psi_k|$ into pure states
\cite{leoneStabilizerEntropiesAre2024}.
Equation~(\ref{eq:SRE2}) is not that extension: it is the \emph{raw} functional, obtained by evaluating the pure-state expression directly on $\rho$.
A general partial-trace monotonicity theorem for Eq.~(\ref{eq:SRE2}) is not available, so neither $\mathcal M^{(2)}(\rho_{ij})\leq\mathcal M^{(2)}(\rho)$ nor $\cM_{i,j}\geq0$ is assumed here; Sec.~V exhibits a state for which $\cM_{i,j}<0$.
The raw functional is moreover not faithful: Sec.~\ref{sec:free} shows that it is nonzero on the two-site marginals studied here even though they are mixtures of stabilizer states and carry no magic at all.
When the goal is to certify nonstabilizerness as a resource rather than to diagnose how it is distributed in space, rigorous mixed-state quantities are the appropriate tool: the robustness of magic, evaluated for spin chains by quantum Monte Carlo tomography
\cite{timsinaRobustnessNonstabilizernessQuantum2025},
and efficiently testable witnesses of mixed-state magic
\cite{haugEfficientWitnessingTesting2026}.

\section{Exact two-site tomography}

We consider number-conserving spinless fermions and order the two-mode Fock basis as
\begin{equation}
\left\{
|0\rangle,\;
c_j^\dagger|0\rangle,\;
c_i^\dagger|0\rangle,\;
c_i^\dagger c_j^\dagger|0\rangle
\right\}.
\end{equation}
Fermion parity forbids matrix elements between the even sector $\{|0\rangle,c_i^\dagger c_j^\dagger|0\rangle\}$ and the odd sector $\{c_j^\dagger|0\rangle,c_i^\dagger|0\rangle\}$, and particle-number conservation additionally removes the anomalous element $\langle c_jc_i\rangle$ linking $|0\rangle$ and $c_i^\dagger c_j^\dagger|0\rangle$.
Five of the fifteen real parameters of a general two-mode density matrix therefore survive; in a pairing state the anomalous element would have to be measured as well.
Direct evaluation of the remaining matrix elements gives
\begin{equation}
\rho_{ij}=
\begin{pmatrix}
1-\langle n_i\rangle-\langle n_j\rangle+\langle n_i n_j\rangle&0&0&0\\
0&\langle n_j\rangle-\langle n_i n_j\rangle&
\langle c_i^\dagger c_j\rangle&0\\
0&\langle c_j^\dagger c_i\rangle&
\langle n_i\rangle-\langle n_i n_j\rangle&0\\
0&0&0&\langle n_i n_j\rangle
\end{pmatrix}.
\label{eq:rho}
\end{equation}
Thus two local densities, one complex Green function, and one density-density correlator are a complete tomography set.
This is the practical reason the estimator scales: all four inputs are standard observables in DMRG and DQMC, and no Pauli- or Majorana-string sampling is needed.

\subsection{Closed form in terms of physical correlators}

Define
\begin{equation}
m_i=1-2\langle n_i\rangle,\qquad
m_j=1-2\langle n_j\rangle,\qquad
\delta_{ij}=\langle n_i n_j\rangle-\langle n_i\rangle\langle n_j\rangle,
\end{equation}
and write $g_{ij}=\langle c_i^\dagger c_j\rangle=g_R+\text{i} g_I$.
The fourth Pauli moment entering Eq.~(\ref{eq:SRE2}) is
\begin{align}
A_4(i,j)
&\equiv\frac14\sum_{P\in\mathcal P_2}
|\Tr(P\rho_{ij})|^4 \nonumber\\
&=\frac14\big[
1+m_i^4+m_j^4+m_i^4m_j^4
+32g_R^4+32g_I^4+256\delta_{ij}^4
\nonumber\\
&\hspace{2.4cm}
+256m_i m_j\delta_{ij}^3
+96m_i^2m_j^2\delta_{ij}^2
+16m_i^3m_j^3\delta_{ij}
\big],
\label{eq:A4}
\end{align}
while the purity is
\begin{equation}
A_2(i,j)\equiv\Tr\rho_{ij}^2
=\frac{(1+m_i^2)(1+m_j^2)}4
+2|g_{ij}|^2+2m_i m_j\delta_{ij}+4\delta_{ij}^2.
\label{eq:A2}
\end{equation}
The two-site functional is simply
$\mathcal M^{(2)}(\rho_{ij})=\ln[A_2(i,j)/A_4(i,j)]$.
For a single site,
\begin{equation}
\mathcal M^{(2)}(\rho_i)
=\ln\frac{1+m_i^2}{1+m_i^4}.
\label{eq:single}
\end{equation}
Equations~(\ref{eq:A4})--(\ref{eq:single}) give the exact connected observable for arbitrary filling and nonuniform local density.

At half filling, $m_i=m_j=0$, and the result becomes especially compact:
\begin{equation}
\cM_{i,j}=
\ln\frac{1+8|g_{ij}|^2+16\delta_{ij}^2}
{1+32(g_R^4+g_I^4)+256\delta_{ij}^4}.
\label{eq:halfexact}
\end{equation}
For correlations small compared with unity,
\begin{equation}
\cM_{i,j}=8|g_{ij}|^2+16\delta_{ij}^2
+O(g_{ij}^4,g_{ij}^2\delta_{ij}^2,\delta_{ij}^4).
\label{eq:halflong}
\end{equation}
This expression supplies the scaling connection used in the Letter.
It also makes clear that the correlator is nonlinear in conventional correlations rather than an independent two-point expectation value.

\subsection{The two-site marginals are stabilizer mixtures}
\label{sec:free}

Equation~(\ref{eq:SRE2}) is nonzero on states with no magic whatsoever, and this is the generic situation here rather than an exception.
Because $\rho_{ij}$ is block diagonal, it is a mixture of stabilizer states as soon as its $\{|01\rangle,|10\rangle\}$ block, viewed as a qubit, lies inside the stabilizer octahedron $|x|+|y|+|z|\leq1$.
Writing $b=\langle n_j\rangle-\langle n_in_j\rangle$ and $c=\langle n_i\rangle-\langle n_in_j\rangle$ for the two diagonal entries of that block, the condition reads
\begin{equation}
2|g_R|+2|g_I|+|b-c|\;\leq\;b+c .
\label{eq:octahedron}
\end{equation}
At half filling with real $g_{ij}$ it reduces to $|g_{ij}|\leq\tfrac14-\delta_{ij}$, which is precisely the positivity condition on $\rho_{ij}$.
Every physical $\rho_{ij}$ is then a stabilizer mixture, as the explicit decomposition
\begin{equation}
\rho_{ij}=
\Big(\tfrac14+\delta_{ij}\Big)\big(|00\rangle\langle00|+|11\rangle\langle11|\big)
+\Big(\tfrac14-\delta_{ij}+g_{ij}\Big)|\Psi^+\rangle\langle\Psi^+|
+\Big(\tfrac14-\delta_{ij}-g_{ij}\Big)|\Psi^-\rangle\langle\Psi^-|
\label{eq:stabdecomp}
\end{equation}
shows, with $|\Psi^\pm\rangle=(|01\rangle\pm|10\rangle)/\sqrt2$ and all four weights non-negative.
We verified Eqs.~(\ref{eq:octahedron})--(\ref{eq:stabdecomp}) by solving the stabilizer-decomposition linear program over all $60$ two-qubit stabilizer states: no physical $\rho_{ij}$ at half filling carries magic, and away from half filling the linear program agrees with Eq.~(\ref{eq:octahedron}) in every instance tested.

This covers every calculation reported here.
Particle--hole symmetry is exact in the auxiliary-field ensemble, so $\langle n_i\rangle=\tfrac12$ throughout the honeycomb study at both zero and finite temperature; the same holds in the one-dimensional Luttinger liquid and across the peak region.
In the symmetry-broken CDW phase the densities are staggered, but $g_{ij}$ is exponentially small at the separations shown and Eq.~(\ref{eq:octahedron}) remains satisfied.
The Laughlin orbital pairs of Sec.~V are diagonal and therefore free as well.
Consequently $\cM_{i,j}$ must be read as a Pauli-moment correlation functional; it is not a local magic density, and the nonzero values reported in the Letter do not certify nonstabilizerness.

Two consequences deserve emphasis.
First, the construction is frame dependent in a controlled way: under $c_i\mapsto e^{\text{i}\theta}c_i$ the combination $g_R^4+g_I^4$ in Eq.~(\ref{eq:halfexact}) changes, and Eq.~(\ref{eq:octahedron}) shows that a generic phase can even push $\rho_{ij}$ out of the stabilizer polytope.
Any magic seen this way is an artifact of the local frame; the real-hopping models studied here sit at exactly zero for every $\theta$ that is a multiple of $\pi/2$.
Second, the global rank-2 SRE obeys a volume law (Sec.~IV), while every two-site marginal is free.
Whatever magic these fermionic states possess is therefore intrinsically nonlocal, invisible to any two-site probe, which also explains why the orbital plateau of Sec.~V cannot by itself signal topological order.

\section{Numerical calculations and scaling analyses}

We study
\begin{equation}
H=-t\sum_{\langle i,j\rangle}
(c_i^\dagger c_j+c_j^\dagger c_i)
+V\sum_{\langle i,j\rangle}
\left(n_i-\frac12\right)\left(n_j-\frac12\right)
\label{eq:tV}
\end{equation}
at half filling and set $t=1$, so all energies and temperatures are quoted in units of the hopping.

\subsection{One-dimensional DMRG and BKT drift}

The chain calculations use matrix-product-state DMRG with open boundaries, as implemented in the Julia version of the ITensor library
\cite{whiteDensityMatrixFormulation1992,itensor,itensor-r0.3}.
The reference site is chosen near the center to reduce boundary effects.
Figure~1(b) of the Letter uses $L=400$, Fig.~1(c) uses $L=900$, and the pseudocritical sequence in Fig.~1(d) extends through $L=900$.
Convergence is controlled by a truncation-error threshold of $10^{-12}$ with the bond dimension capped at $3000$.
The cap is never reached: the truncation criterion is always met first, and even in the Luttinger-liquid phase at $L=900$ the maximum bond dimension remains below $800$.

In the Luttinger liquid and throughout the peak region, the measured local densities equal $1/2$ to numerical precision, so Eq.~(\ref{eq:halflong}) applies; its inputs decay algebraically and therefore so does $\cM$.
Deep in the gapped charge-density-wave (CDW) regime, by contrast, the ground state selected by DMRG breaks particle--hole symmetry spontaneously, the local densities are staggered, and we evaluate the general expressions~(\ref{eq:A4})--(\ref{eq:single}).
Expanding these for small $g_{ij}$ and $\delta_{ij}$ at fixed $m_i$ gives
\begin{equation}
\cM_{i,j}=c_1\delta_{ij}
+\frac{8|g_{ij}|^2}{(1+m_i^2)(1+m_j^2)}
+O(\delta_{ij}^2),
\qquad
c_1=\frac{8m_im_j}{(1+m_i^2)(1+m_j^2)}
-\frac{16m_i^3m_j^3}{(1+m_i^4)(1+m_j^4)},
\label{eq:brokenexpansion}
\end{equation}
the constant terms cancelling identically against the single-site subtractions for any $m_i$ and $m_j$.
The leading term is linear rather than quadratic in $\delta_{ij}$, so in this phase $\cM$ decays with the density correlation length itself; both connected correlators decay exponentially and so does $\cM$.
Had the symmetry instead been enforced, $\delta_{ij}$ would saturate at $(-1)^{i-j}\mu^2$, with $\langle n_i\rangle=\tfrac12+(-1)^i\mu$ the staggered density of a single ordered branch, and $\cM$ would approach the plateau $\ln[(1+16\mu^4)/(1+256\mu^8)]$; both behaviours identify the ordered phase.
The fits shown in Fig.~1(b) of the Letter are
\begin{equation}
\cM_{i,i+r}\propto r^{-1.91}\quad(V=1),\qquad
\cM_{i,i+r}\propto e^{-r/4.36}\quad(V=4).
\end{equation}
These exponents and length scales are empirical fits to the stabilizer-R\'enyi correlator; they are not separately asserted to equal the Green-function exponents or correlation lengths.

For $V>2$, write $V/2=\cosh\lambda$.
The Bethe-ansatz BKT scale has the leading form
\begin{equation}
\xi=\xi_0\exp\!\left(\frac{\pi^2}{2\lambda}\right),
\qquad
\lambda=\operatorname{arccosh}(V/2)
\simeq\sqrt{V-2}.
\label{eq:xiBKT}
\end{equation}
Equating $\xi$ with $L$ yields
\begin{equation}
V_c(L)=2+\frac{\pi^4}{4[\ln L-\ln L_0]^2},
\label{eq:BKTdrift}
\end{equation}
where $L_0$ absorbs $\xi_0$ and subleading corrections.
Fitting sizes $L>100$ gives $L_0\simeq1.7$, as reported in the Letter.

\subsection{Honeycomb-lattice DQMC}

For the honeycomb lattice we employ auxiliary-field DQMC
\cite{blankenbeclerMonteCarloCalculations1981,scalapinoMonteCarloCalculations1981,assaadWorldlineDeterminantalQuantum2008}.
A Trotter--Suzuki decomposition and Hubbard--Stratonovich transformation express the interacting density matrix as an ensemble of fermion-bilinear evolution operators.
At half filling, the nearest-neighbor spinless model admits a sign-problem-free formulation
\cite{Huffman2014,Li2015solving,Wang2015split}.
For each auxiliary-field configuration we measure the equal-time Green function and density correlations, average these physical observables over the Monte Carlo ensemble, construct Eq.~(\ref{eq:rho}), and only then evaluate Eq.~(\ref{eq:connected}).
This order is essential: averaging the nonlinear SRE of configuration-resolved Gaussian operators would not give the SRE of the interacting density matrix.
Statistical uncertainties are propagated through the nonlinear tomography formula using a jackknife analysis over Monte Carlo bins.
The zero-temperature data are obtained with projector DQMC rather than by approximating the ground state using finite-temperature DQMC at a sufficiently low temperature.
Starting from a trial Slater determinant $|\Psi_T\rangle$ having a nonzero overlap with the ground state, an observable $\hat O$ is evaluated as
\begin{equation}
\langle\hat O\rangle=
\frac{\langle\Psi_T|e^{-\Theta\hat H}\hat O e^{-\Theta\hat H}|\Psi_T\rangle}
{\langle\Psi_T|e^{-2\Theta\hat H}|\Psi_T\rangle}.
\label{eq:PQMCprojection}
\end{equation}
We choose $|\Psi_T\rangle$ as the half-filled ground state of the noninteracting Hamiltonian, adding weak random nearest-neighbor hoppings only to lift the finite-size degeneracy at the Dirac points.
The total projection length $2\Theta$ is divided into $L_\tau$ slices, $2\Theta=L_\tau\Delta\tau$, and a symmetric Trotter decomposition is applied on every slice,
\begin{equation}
e^{-\Delta\tau\hat H}\simeq e^{-\Delta\tau\hat H_0/2}e^{-\Delta\tau\hat H_I}e^{-\Delta\tau\hat H_0/2},
\end{equation}
whose leading accumulated discretization error is of order $(\Delta\tau)^2$.
A discrete Hubbard--Stratonovich field on each nearest-neighbor bond and time slice decouples $\hat H_I$ in the hopping channel.
The fermions can then be integrated out, leaving a determinant weight for each auxiliary-field configuration.
We sample this weight with local Metropolis updates, stabilize the products of single-particle propagators, and measure equal-time observables at the midpoint of the projection.
With $t=1$, the honeycomb simulations use $\Delta\tau t=0.05$ and a size-dependent total projection length $2\Theta t=L+12$, following Ref.~\cite{wangfohong2026}.

For the finite-temperature calculations, we instead use conventional finite-temperature DQMC and evaluate the thermal trace $Z=\Tr e^{-\beta\hat H}$, with $\beta=L_\tau\Delta\tau$.
Thus the thermal calculation requires neither a trial Slater determinant nor random hopping perturbations to lift the finite-size degeneracy at the Dirac points.
Away from the thermal critical region we use $\Delta\tau t=0.05$; close to the transition we reduce the time step to $\Delta\tau t=0.01$ to better control the Trotter error.

\paragraph{Quantum transition.}
For maximally separated sites, the CDW correlator scales as
\begin{equation}
\delta_{i,i+\bm r_{\max}}\sim L^{-(d+z-2+\eta)}
=L^{-(1+\eta)}
\end{equation}
for $d=2$ and $z=1$.
The fermion Green function scales as
$g_{i,i+\bm r_{\max}}\sim L^{-(2+\eta_\psi)}$.
Which of the two channels in Eq.~(\ref{eq:halflong}) dominates is a comparison of scaling dimensions.
In the $D=d+z=3$ dimensional conformal field theory describing the transition, the order-parameter and fermion operators have
\begin{equation}
\Delta_\phi=\frac{1+\eta}{2},
\qquad
\Delta_\psi=\frac{2+\eta_\psi}{2},
\label{eq:dimensions}
\end{equation}
and the equal-time correlators above are $r^{-2\Delta_\phi}$ and $r^{-2\Delta_\psi}$.
The unitarity bounds for scalar and spin-$\tfrac12$ operators, $\Delta_\phi\geq(D-2)/2=1/2$ and $\Delta_\psi\geq(D-1)/2=1$, are identically the statements $\eta\geq0$ and $\eta_\psi\geq0$, with equality only for free fields.
The Green function therefore never decays more slowly than $L^{-2}$, and the density channel leads whenever $\Delta_\phi<1$, that is, $\eta<1$.

This inequality is not accidental in the present class.
In $D=3$, $\eta=1$ corresponds to $\Delta_\phi=1$ and hence to an order-parameter propagator $D_\phi(p)\sim1/|p|$, which is precisely what gapless Dirac fermions generate by themselves: the particle--hole bubble gives $\Pi(p)\propto|p|^{D-2}=|p|$, and at $N\to\infty$ this bubble is the entire inverse propagator.
Thus $\eta=1$ is the large-$N$ saturation value of the Gross--Neveu--Yukawa class, approached from below, and $1/N$ corrections reduce it to $\eta\simeq0.5$ at the $N=4$ chiral Ising fixed point relevant here, with $\eta_\psi\simeq0.1$
\cite{Phys.Rev.B2018Ihrig,graceyAnomalousDimensionsCritical2025,Phys.Lett.B1992Gracey,J.HighEnerg.Phys.2023Erramilli}.
The separation is comfortable, $1+\eta\simeq1.5$ against $2+\eta_\psi\simeq2.1$, and
\begin{equation}
\cM_{i,i+\bm r_{\max}}\sim
\delta_{i,i+\bm r_{\max}}^2\sim L^{-2(1+\eta)}.
\label{eq:QCPscale}
\end{equation}
Were the ordering reversed, Eq.~(\ref{eq:halflong}) would instead predict $\cM_{i,i+\bm r_{\max}}\sim L^{-2(2+\eta_\psi)}$; the construction itself is unchanged.
We emphasize that this argument uses Lorentz invariance ($z=1$) and does not extend unmodified to metallic quantum critical points with a Fermi surface.

\begin{figure}[t]
\centering
\includegraphics[width=0.5\textwidth]{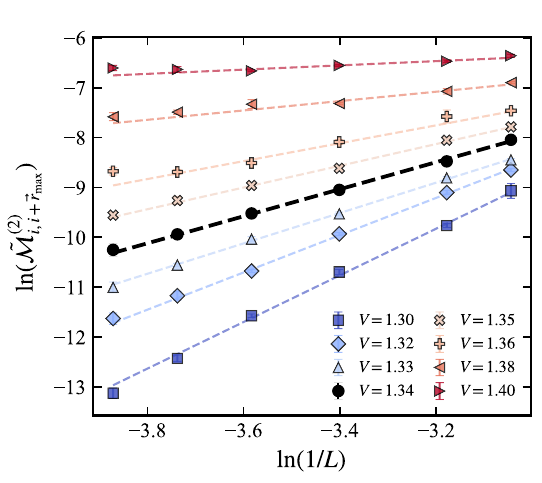}
\caption{\textbf{Longest-distance stabilizer-R\'enyi correlator across the honeycomb transition.}
DQMC results for $\cM_{i,i+\bm r_{\max}}$ on honeycomb lattices up to $L=48$ at zero temperature.
The dashed line at $V=1.34$ shows critical power-law behavior.
Error bars show DQMC statistical uncertainties.}
\label{fig:rawdata}
\end{figure}

The dominant channel also changes away from criticality, which accounts for the trends on either side of $V_c$.
In the Dirac semimetal, Wick's theorem gives $\delta_{i,i+\bm r_{\max}}\sim|g_{i,i+\bm r_{\max}}|^2\sim L^{-4}$ while $g_{i,i+\bm r_{\max}}\sim L^{-2}$, so the Green-function term leads and $\cM_{i,i+\bm r_{\max}}\sim L^{-4}$.
At the critical point the density term takes over with the slower decay $L^{-2(1+\eta)}\simeq L^{-3}$.
In the ordered phase $\delta_{i,i+\bm r_{\max}}$ approaches a nonzero constant set by the CDW order parameter, and $\cM_{i,i+\bm r_{\max}}$ saturates.
Figure~\ref{fig:rawdata} shows the raw data for the longest-distance stabilizer-R\'enyi correlator.
Around $V=1.34$ the signal follows a power law over the available sizes, consistent with Eq.~(\ref{eq:QCPscale}), while the semimetallic and ordered trends produce visible departures on either side.

We use the CDW correlation ratio
\begin{equation}
R=1-\frac{C(\bm q_{\min})}{C(\bm0)},\qquad
\bm q_{\min}=(2\pi/L,0),
\label{eq:ratio}
\end{equation}
as a dimensionless scaling coordinate and apply the kernel method of
Ref.~\cite{haradaKernelMethodCorrections2015}.
The data for
$L=21,24,30,36,42,48$ and $V\leq1.36$ yield the collapse in Fig.~2(b) of the Letter with $\eta=0.423(76)$.
Couplings $V>1.36$ are not used in the collapse because ordered-phase corrections are visibly large.

\paragraph{Thermal transition.}
At fixed interaction in the CDW regime, the finite-temperature transition is in the two-dimensional Ising universality class.
With $\eta_{\mathrm I}=1/4$ and $\nu_{\mathrm I}=1$, Eq.~(\ref{eq:halflong}) gives
\begin{equation}
\cM(T,L)=L^{-1/2}
\mathcal G\!\left(\frac{T-T_c}{T_c}L\right).
\label{eq:thermalcollapse}
\end{equation}
Figure~2(c) of the Letter shows the resulting collapse for
$L=12,18,24,30$ at fixed $V=2.0$ in the CDW phase.
The kernel-based data-collapse analysis \cite{haradaKernelMethodCorrections2015} yields a critical temperature of $T_c/t=0.4740(4)$.

\section{Global SRE estimator and its sampling bottleneck}

For completeness, we summarize the global fermionic SRE calculation and its bottleneck.
Introduce $2N$ Majorana operators
\begin{equation}
\gamma_{2i-1}=c_i^\dagger+c_i,\qquad
\gamma_{2i}=\text{i}(c_i^\dagger-c_i),
\end{equation}
which obey $\{\gamma_\mu,\gamma_\nu\}=2\delta_{\mu\nu}$.
The Jordan--Wigner transformation gives a bijection, up to an irrelevant phase, between Pauli strings and Majorana monomials
\begin{equation}
\gamma^{\bm x}=\gamma_1^{x_1}\gamma_2^{x_2}\cdots
\gamma_{2N}^{x_{2N}},\qquad
\bm x\in\{0,1\}^{2N}.
\end{equation}
Therefore
\begin{equation}
\sum_{P\in\mathcal P_N}|\Tr(P\rho)|^{2\alpha}
=\sum_{\bm x}|\Tr(\gamma^{\bm x}\rho)|^{2\alpha}.
\label{eq:PauliMajorana}
\end{equation}

For a fermionic Gaussian state $\rho_G$, Wick's theorem gives
\begin{equation}
\Tr(\gamma^{\bm x}\rho_G)
=(-\text{i})^{|\bm x|/2}\Pf(\Gamma_G|_{\bm x}),
\label{eq:pf}
\end{equation}
for even $|\bm x|$; odd-parity expectations vanish.
We use the real antisymmetric covariance-matrix convention
\begin{equation}
(\Gamma_G)_{\mu\nu}
=\frac{\text{i}}{2}\Tr\!\left(\rho_G[\gamma_\mu,\gamma_\nu]\right).
\label{eq:covariance}
\end{equation}
The phase in Eq.~(\ref{eq:pf}) follows from this convention and drops out of the SRE moments.
Here $\Gamma_G|_{\bm x}$ is restricted to the indices selected by $\bm x$.
For a pure Gaussian state,
\begin{equation}
\pi_G(\bm x)=
\frac{\det(\Gamma_G|_{\bm x})}{\det(1+\Gamma_G)}
\label{eq:probability}
\end{equation}
is normalized, and the SRE can be sampled from its R\'enyi moment
\cite{colluraQuantumMagicFermionic2025}.

An interacting DQMC density matrix is an auxiliary-field ensemble,
\begin{equation}
\rho=\sum_{\bm s}p_{\bm s}\rho_{\bm s},
\end{equation}
where each $\rho_{\bm s}$ is Gaussian.
Hence
\begin{equation}
\Tr(\gamma^{\bm x}\rho)
=(-\text{i})^{|\bm x|/2}\sum_{\bm s}p_{\bm s}\Pf(\Gamma_{\bm s}|_{\bm x}).
\label{eq:nested}
\end{equation}
Every Majorana configuration in the outer SRE sum thus requires an inner auxiliary-field estimate.

The nesting originates in two features: the SRE is exponentially small in the system size, and the string amplitudes are sign indefinite.
Because the phase in Eq.~(\ref{eq:pf}) drops out of the moments, the fourth moment takes the four-replica form
\begin{equation}
\left|\Tr(\gamma^{\bm x}\rho)\right|^{4}
=\sum_{\bm s_1,\ldots,\bm s_4}
\prod_{k=1}^{4}p_{\bm s_k}\Pf(\Gamma_{\bm s_k}|_{\bm x}),
\label{eq:replica}
\end{equation}
so a flat Monte Carlo over the joint space $(\bm x,\bm s_1,\ldots,\bm s_4)$ would avoid nesting altogether.
Its summand splits into a manifestly positive factor $\prod_{k}p_{\bm s_k}$ and a sign-indefinite factor $\prod_{k}\Pf(\Gamma_{\bm s_k}|_{\bm x})$.
Sampling the auxiliary fields from the first and measuring the second is unbiased and uses only positive weights, but it does not scale: the global SRE obeys a volume law, so the measured quantity is exponentially small in $N$, while the individual Pfaffians---minors selected by system-size Majorana strings---are broadly distributed in both magnitude and sign.
Controlling an exponentially small mean requires importance sampling of the $2^{2N}$ string configurations, and hence a non-negative weight in $\bm x$.
Taking the modulus of the summand supplies one, at the price of carrying its sign in the estimator, so the cancellation returns as a conventional sign problem even though the model is sign-problem-free for conventional observables, $p_{\bm s}\geq0$.
The only manifestly non-negative string weight available is
$|\Tr(\gamma^{\bm x}\rho)|^{4}=[\sum_{\bm s}p_{\bm s}\Pf(\Gamma_{\bm s}|_{\bm x})]^{4}$,
in which the field average---itself real but not sign-definite, Eq.~(\ref{eq:nested})---is taken before the fourth power; evaluating it for every string is precisely the nested scheme.
The sign fluctuations do not disappear there: they resurface as the variance of the inner estimate, which demands exponentially many field configurations once the averaged amplitude becomes small for long strings.
Updating two local Pauli factors produces a nonlocal Majorana update through the Jordan--Wigner string and reduces autocorrelation, but it does not remove this obstruction.
For spin models the corresponding weights can be made positive, and the $\alpha$-SRE is then computable as a ratio of partition functions without nesting
\cite{dingEvaluatingManybodyStabilizer2025}.
The obstruction described here is specific to the Majorana-string representation of Eq.~(\ref{eq:PauliMajorana}); we do not claim that no positive-weight estimator of the global fermionic SRE exists.

\begin{figure}[t]
\centering
\includegraphics[width=0.62\textwidth]{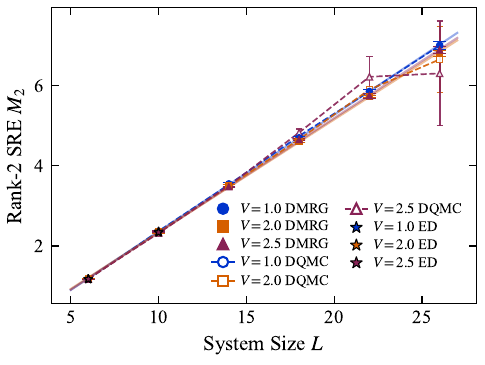}
\caption{\textbf{Global rank-2 SRE benchmark.}
Global $M_2$ for periodic half-filled $t$-$V$ chains at $V=1,2,2.5$.
DMRG, DQMC, and exact diagonalization agree where the DQMC estimator is resolved and show volume-law growth.
At larger sizes, especially for $V=2.5$, the DQMC uncertainty grows rapidly.
This benchmark demonstrates correctness at small sizes and motivates the two-site construction.}
\label{fig:global}
\end{figure}

Figure~\ref{fig:global} benchmarks the method against DMRG and exact diagonalization.
It demonstrates correctness at accessible sizes and volume-law growth, but also the rapidly increasing DQMC uncertainty that the sign structure above implies.
This result documents a limitation rather than a scalable capability.

\section{Exploratory Laughlin-orbital calculation}

Finally, we record an exploratory Laughlin-orbital calculation, with its scope stated explicitly.
For the fermionic $\nu=1/m$ Laughlin state on the sphere,
$N_\phi=m(N-1)$ and the lowest Landau level contains $N_\phi+1$ orbitals.
In Schwinger-boson coordinates,
\begin{equation}
|\Psi_m\rangle\propto
\prod_{a<b}(u_a v_b-v_a u_b)^m|0\rangle
\label{eq:laughlin}
\end{equation}
\cite{laughlinAnomalousQuantumHall1983,haldaneFractionalQuantizationHall1983,haqueEntanglementEntropyFermionic2007}.
Rotational symmetry makes the single-orbital occupation uniform,
\begin{equation}
\langle n_l\rangle=\nu_{\mathrm{eff}}
=\frac{N}{m(N-1)+1}.
\end{equation}
For two distinct angular-momentum orbitals $(0,j)$,
$\langle c_0^\dagger c_j\rangle=0$, so their reduced density matrix is diagonal and determined by
$x=P_{11}(0,j)=\langle n_0n_j\rangle$.

The factor in Eq.~(\ref{eq:laughlin}) implies an exact orbital exclusion window.
If one particle occupies orbital $0$, every other particle carries at least $m$ units of the complementary Schwinger-boson power.
Thus
\begin{equation}
P_{11}(0,j)=0,\qquad j<m.
\label{eq:hole}
\end{equation}
Within this window,
\begin{equation}
\rho_{0j}=\operatorname{diag}
(1-2\nu_{\mathrm{eff}},\nu_{\mathrm{eff}},
\nu_{\mathrm{eff}},0),
\end{equation}
and the only nonzero Pauli expectations are
\begin{equation}
\langle II\rangle=1,\quad
\langle ZI\rangle=\langle IZ\rangle=1-2\nu_{\mathrm{eff}},\quad
\langle ZZ\rangle=1-4\nu_{\mathrm{eff}}.
\end{equation}
Using Eqs.~(\ref{eq:SRE2ratio}) and (\ref{eq:connected}) gives the plateau
\begin{align}
\tilde{\mathcal M}^{(2)}_{\mathrm{plateau}}
={}&
\ln\frac{1+2(1-2\nu_{\mathrm{eff}})^2+(1-4\nu_{\mathrm{eff}})^2}
{1+2(1-2\nu_{\mathrm{eff}})^4+(1-4\nu_{\mathrm{eff}})^4}
\nonumber\\
&-2\ln\frac{1+(1-2\nu_{\mathrm{eff}})^2}
{1+(1-2\nu_{\mathrm{eff}})^4}.
\label{eq:plateau}
\end{align}
For $m=3$, $N=9$, this gives $\nu_{\mathrm{eff}}=9/25$ and
$\tilde{\mathcal M}^{(2)}_{\mathrm{plateau}}\simeq0.113$.
For $m=5$, $N=6$, it gives
$\nu_{\mathrm{eff}}=3/13$ and
$\tilde{\mathcal M}^{(2)}_{\mathrm{plateau}}\simeq-0.042$.
At fixed $\nu_{\mathrm{eff}}$, the stationary point as a function of
$x=P_{11}$ is
\begin{equation}
x_c=\nu_{\mathrm{eff}}-\frac14.
\end{equation}

\begin{figure}[t]
\centering
\begin{minipage}{0.47\textwidth}
\centering
\includegraphics[width=\linewidth]{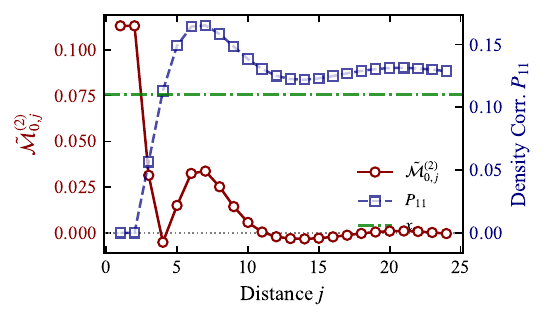}\\[-2mm]
\textbf{(a)}
\end{minipage}
\hfill
\begin{minipage}{0.47\textwidth}
\centering
\includegraphics[width=0.82\linewidth]{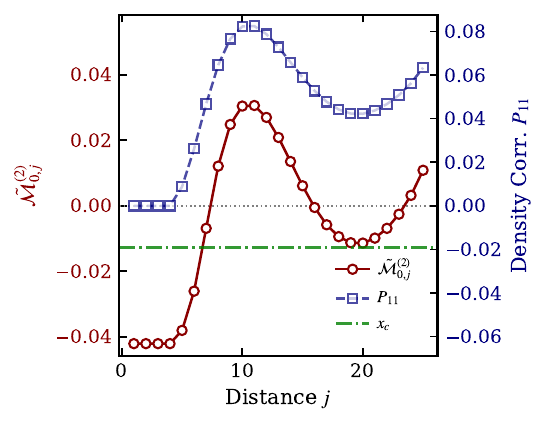}\\[-2mm]
\textbf{(b)}
\end{minipage}
\caption{\textbf{Orbital exclusion structure in finite Laughlin states.}
The connected SRE diagnostic $\cM_{0,j}$ and joint occupation
$P_{11}(0,j)$ for (a) $\nu=1/3$, $N=9$ and
(b) $\nu=1/5$, $N=6$.
The flat regions for $j<m$ follow from the exact exclusion in Eq.~(\ref{eq:hole}); $x_c$ marks the stationary value of the connected SRE as a function of $P_{11}$.}
\label{fig:laughlin}
\end{figure}

Figure~\ref{fig:laughlin} verifies the exclusion formula and shows how the nonlinear diagnostic follows orbital occupation correlations outside the hole.
Three qualifications delimit the scope of this result.
First, the orbital index $j$ is not a real-space separation, and the construction depends on the chosen one-particle basis and reference orbital.
Second, the negative $m=5$ plateau explicitly illustrates that $\cM$ is not a mutual-information-like nonnegative monotone.
Third, a two-orbital exclusion pattern is local data and by itself cannot establish long-range topological order.
The result is therefore an instructive application and an open direction for higher-order or genuinely nonlocal magic diagnostics, not evidence that Eq.~(\ref{eq:connected}) detects topological order.

\bibliography{ref}